\documentclass[12pt,letterpaper]{article}
\pdfoutput=1
\usepackage{jcappub}
\usepackage{bm}
\usepackage{amsmath,amssymb}
\usepackage{subfig}
\usepackage{caption}

\newcommand{\Beq}{\begin{equation}\begin{aligned}}
\newcommand{\Eeq}{\end{aligned}\end{equation}}

\usepackage{graphicx, color}
\usepackage{dcolumn}
\usepackage{bm}
\usepackage[mathlines]{lineno}
\usepackage{amsmath}
\usepackage{amssymb}
\usepackage[numbers,sort&compress]{natbib}
\usepackage{bm}
\usepackage{float}
\usepackage{mathrsfs}
\usepackage{lipsum}
\usepackage{hyperref}
\usepackage[normalem]{ulem}
\hypersetup{
    colorlinks=true,       
    linkcolor=rp,          
    citecolor=green2,        
    filecolor=magenta,      
    urlcolor=green2           
}

\usepackage[all]{hypcap} 
\usepackage{cleveref}

\newcommand{\beq}{\begin{align}}
\newcommand{\eeq}{\end\begin{align}}

\def\lap{\lower.5ex\hbox{$\; \buildrel < \over \sim \;$}}
\def\gap{\lower.5ex\hbox{$\; \buildrel > \over \sim \;$}}

\def\be{\begin{align}}
\def\ee{\end\begin{align}}
\def\ba{\begin{eqnarray}}
\def\ea{\end{eqnarray}}

\def\b{\boldsymbol}
\def\bk{\b k}

\def\bx{\b x}

\def\inter{{\rm int}}

\def\kpc{\textrm{kpc}}

\definecolor{rp}{cmyk}{0.2, 1, 0.6, 0}
\definecolor{rp}{cmyk}{0.2, 1, 0.6, 0}
\definecolor{green2}{cmyk}{0.27, 0, 1, 0.52}

\newcommand{\bPsi}{{\bm{\mathsf{\Psi}}}}
\newcommand{\bepsilon}{{\bm{\mathsf{\epsilon}}}}
\newcommand{\bS}{{\bm{S}}}
\newcommand{\bW}{{\bm{W}}}

\newcommand{\mpl}{m_{\mathrm{pl}}}

\title{\huge Small-scale structure in  vector dark matter}

\author[a]{Mustafa A. Amin,}
\author[a]{Mudit Jain,}
\author[a,b]{Rohith Karur,}
\author[c]{Philip Mocz}

\affiliation[a]{Department of Physics and Astronomy, Rice University, Houston, Texas 77005, U.S.A.}
\affiliation[b]{Department of Physics, University of California, Berkeley, California 94720, U.S.A.}
\affiliation[c]{Department of Astrophysical Sciences, Princeton University, Princeton, NJ 08544, U.S.A}

\emailAdd{mustafa.a.amin@rice.edu}
\emailAdd{mudit.jain@rice.edu}
\emailAdd{r\_karur137@berkeley.edu}
\emailAdd{pmocz@astro.princeton.edu}

\abstract{We investigate the differences in the  small-scale structure of vector dark matter (VDM) and scalar dark matter (SDM) using 3+1 dimensional simulations of single/multicomponent Schr\"{o}dinger-Poisson system.  We find that the amount of wave interference, core-to-halo mass ratio (and its scatter), spin of the core, as well as the shape of the central regions of dark matter halos can distinguish VDM and SDM. Starting with a collection of idealized halos (self-gravitating solitons) as an initial condition, we show that the system dynamically evolves to an approximately spherically symmetric configuration that has a core surrounded by a halo of interference patterns in the mass density. In the vector case, the central soliton in less dense and has a smoother transition to an $r^{-3}$ tail compared to the scalar case. As compared to SDM, wave interference in VDM is $\sim 1/\sqrt{3}$ times smaller, resulting in fewer low and high density regions, and more diffuse granules in the halo. The ratio of VDM core mass to the total halo mass is lower than that in SDM, with a steeper dependence on the total energy of the system and a slightly larger scatter. Finally, we also initiate a study of the evolution of intrinsic spin angular momentum in the VDM case. We see a positive correlation between the total intrinsic spin in the simulation and the spin of the final central core, with significant scatter. We see large intrinsic spin in the  core being possible even with vanishing amounts total angular momentum in the initial conditions (at least instantaneously). Our results point towards the possibility of distinguishing VDM from SDM using astrophysical and terrestrial observations.}

\begin{document}

\maketitle

\section{Introduction}\label{sec:intro}

The identity of particles/fields that make up dark matter remains  uncertain. Observational evidence \cite{Planck:2018vyg} suggests that dark matter is very weakly interacting with the Standard model, is non-relativistic in the contemporary universe, and has clumped efficiently under the influence of gravity (for a historical overview, see \cite{Bertone:2016nfn}). However, the mass and spin of the fundamental quanta of dark matter are not known. The mass of the fundamental particles making up dark matter can range from $\sim 10^{-21}\rm eV$ \cite{Irsic:2017yje} to  $\sim \mpl$. The  bounds are softened further if dark matter is multi-component, or composite \cite{Jacobs:2014yca}. We have no robust constraints on the spin of the particles/fields that make up dark matter either.  

When dark matter is light, $m\ll 10\,\textrm{eV}$, and bosonic, the typical occupation number of the field and the overlap of the particle de Broglie wavelengths is high enough in astrophysical and cosmological settings  that a classical wave description becomes appropriate (as opposed to discrete point-like particle as in CDM)\cite{Hui:2021tkt}.  This ``wave dark matter" results in novel phenomenon resulting from coherence and interference of the effective classical field, common to all wave dynamics. If the mass of the underlying boson is sufficiently small, these wave effects can manifest themselves on macroscopic and even astrophysical scales. Significant effort has been devoted in recent years to study such wave phenomenon, including suppression of structure on small scales \cite{Matos:2000ss,Hu:2000ke}, solitons \cite{Schive:2014dra, Amin:2019ums,Arvanitaki:2019rax}, interference patterns \cite{Schive:2014dra}, turbulence \cite{Mocz:2017wlg}, vortices \cite{Hui:2020hbq}, superradiance \cite{Brito:2015oca} etc. in the context of spin-0 dark matter (scalar dark matter or SDM) in the nonrelativistic limit. See \cite{Marsh:2015wka,Hui:2021tkt,Niemeyer:2019aqm,Ferreira:2020fam} for reviews and references therein. The String Axiverse \cite{Arvanitaki:2009fg,Cicoli:2021gss} as well as the QCD axion and axion-like particles \cite{Wilczek:1977pj,Peccei:1977hh,Preskill:1982cy,Abbott:1982af,Dine:1982ah,Ringwald:2014vqa} provide natural candidates for light, weakly interacting scalars with the above phenomenology.

There is no evidence yet, however, that dark matter is a light scalar field. It is natural to ask about the intrinsic spin of the underlying wave dark matter.  Gravitational production of light (but not ultralight) vector dark matter was explored in detail in \cite{Graham:2015rva} (also see \cite{Ema:2019yrd,Kolb:2020fwh,Ahmed:2020fhc}). Different production mechanisms where the initial energy is stored in a misaligned axion/Higgs  \cite{Agrawal:2018vin,Co:2018lka,Dror:2018pdh,Bastero-Gil:2018uel}, in axion rotations \cite{Co:2021rhi}, or in strings in \cite{Long:2015cza}, and then converted to vector dark matter have also been investigated. In contrast to the gravitational production mechanism, many of these allow for ultralight vector boson masses. Numerical simulations of vector dark matter production in the misaligned axion scenario were carried out in \cite{Agrawal:2018vin}. However, exploring the astrophysical implications of vector wave dark matter in the contemporary universe on nonlinear scales has been restricted to analytic calculations \cite{Adshead:2021kvl,Jain:2021pnk,Blinov:2021axd}.

With this in mind, we carry out 3+1 dimensional simulations of spin-$1$ wave dark matter (which we refer to as vector dark matter or VDM) in the nonrelativistic limit.  In this limit, the dynamics of vector dark matter can be simulated by a multicomponent Schr\"{o}dinger-Poisson system \cite{Adshead:2021kvl,Jain:2021pnk}. This essentially allows us to use the existing numerical framework used for scalar dark matter, but now with three (equal mass) components. We restrict our attention to the mergers of idealized halos, in the form of solitons, to explore how structure formation might proceed on nonlinear scales in VDM. We characterize the core and halo resulting from these mergers, and compare the results with the corresponding SDM case. For identical initial conditions in terms of their initial mass density, the final configurations for SDM and VDM differ from each other. Even if we ignore the knowledge of initial conditions, and simply compare the shape of the final configuration, we are able to distinguish the VDM from the SDM case. We also track the spin density of VDM (which is trivially zero in SDM) for the first time, which could provide another novel handle on VDM when coupled to (for example) the electromagnetic field, or when approaching the relativistic limit \cite{Jain:2021pnk}. Thus, SDM and VDM can be potentially distinguished in an astrophysical setting giving us hope of observationally probing a fundamental degree of freedom (spin) of ultralight dark matter. The key to understanding the differences between SDM and VDM is that the wave-interference effects are smaller in VDM compared to SDM.\footnote{We note that in the non-relativistic limit and {\it without} additional interactions, we cannot distinguish between three scalars with identical masses and a vector field with the same boson mass (although effects related to the production mechanism might hold clues to the underlying nature of the field). The spin discussed here can then be thought of as an isospin \cite{Jain:2021pnk}. Separately, for phenomenology of multiple axion species with different masses in the late universe, see for example, \cite{Arvanitaki:2009fg,Eby:2020eas,Street:2022nib,Guo:2020tla,Cyncynates:2021xzw}.}

The rest of the paper is organized as follows. In Sec.~\ref{sec:model}, we introduce our model for VDM along with its nonrelativistic limit. We also provide an understanding of interference in VDM waves, as well as solitons in VDM. We explore binary soliton mergers in Sec.~\ref{sec:two_merger}, and calculate the fraction of total mass that remains bound in the final soliton. In Sec.~\ref{sec:many_merger}, we consider the merger of $N=\mathcal{O}(10)$ solitons. We compare the results of the merger in VDM and SDM, including core mass, density profiles, size of interference granules, as well as spin angular momentum density. In Sec.~\ref{sec:end}, we briefly discuss observational implications including dynamical heating of stars, cores of dwarf galaxies, and DM substructure. We summarize our main results, as well a future outlook in \ref{sec:summary}. Details of the numerical simulation, as well as some details of our analytic calculations are deferred to the Appendix.
\section{Preliminaries}\label{sec:model}
\subsection{Model and equations of motion}
A (dark) massive spin-$1$ field $W_\mu$ minimally coupled to gravity and without non-gravitational self-interactions is described by the following action:
\begin{align}\label{eq:rel_action}
    S &= \int\mathrm{d}^4x\,\sqrt{-g}\Bigl[-\frac{1}{4}g^{\mu\alpha}g^{\nu\beta}\,\mathcal{G}_{\mu\nu}\mathcal{G}_{\alpha\beta} + \frac{1}{2}\frac{m^2c^2}{\hbar^2}\,g^{\mu\nu}W_{\mu}W_{\nu} + \frac{c^3}{16\pi G}R + ...\Bigr],
\end{align}
where $\mathcal{G}_{\mu\nu}=\partial_\mu W_\nu-\partial_\nu W_\mu$. The `$...$' in \eqref{eq:rel_action} represents the Standard Model Lagrangian and other possible dark sector(s). Here, $m$ is the mass of the vector boson. We can represent the spatial part of the (real-valued) vector field $\bW$ in terms of a complex vector $\bPsi$ as
\begin{align}
\label{nonrelativistic_expansion}
&\bW(t,\bx) \equiv \frac{\hbar}{\sqrt{2mc}}\Re\left[\bPsi(t,\b x) e^{-imc^2t/\hbar}\right],
\end{align}
where $\bPsi$ has dimensions of $[\textrm{length}]^{-3/2}$. Similarly, $W_0(t,\bx)\equiv\hbar/\sqrt{2mc}\,\Re\left[ \psi_0(t,\b x) e^{-imc^2t/\hbar}\right]$. We are interested in the non-relativistic behaviour of the vector field where the spatial variation in the field is slow compared to the Compton scale $\lambda_m=\hbar/mc$ and we are in the Newtonian gravity regime. We focus on sufficiently subhorizon dynamics, and hence ignore Hubble expansion. In this case, the dynamics are described by the non-relativistic action for the complex vector field $\bPsi$ and the Newtonian gravitational potential $\Phi$: 
\begin{align}\label{eq:nonrel_action}
    \mathcal{S}_{nr} &= \int \mathrm dt \mathrm d^3x\,\Biggl[\frac{i\hbar}{2}\bPsi^{\dagger}\dot{\bPsi} + \mathrm{c.c.} - \frac{\hbar^2}{2m} \nabla\bPsi^{\dagger}\cdot\nabla\bPsi+ \frac{1}{8\pi G}\,\Phi\nabla^2\Phi - m\,\Phi\,\bPsi^{\dagger}\bPsi\Biggr],
\end{align}
and the corresponding multi-component Schr\"{o}dinger-Poisson (SP) system of equations of motion:\footnote{To include the effects of Hubble expansion, simply replace $\nabla\rightarrow \nabla/a$ and $\partial_t\rightarrow \partial_t+3H/2$ where $a$ is the scalefactor and $H=\dot{a}/a$.}
\begin{align}\label{eq:SP_general}
    i\hbar\frac{\partial}{\partial t}\bPsi &= -\frac{\hbar^2}{2m}\nabla^2\bPsi + m\,\Phi\,\bPsi\,,\qquad\qquad
    \nabla^2\Phi = 4\pi Gm\,\bPsi^\dagger \bPsi.
\end{align}
This is our master equation that we work with throughout this work. 
We re-iterate that $\bPsi$ a complex 3-tuple with components $[\bPsi]_i=\psi_i$ with $i=1,2,3$ and $\bPsi^\dagger \bPsi=\sum_{i=1}^3|\psi_i|^2$.  For scalar dark matter, we have a single component field (which leads to the ``usual" Schr\"{o}dinger-Poisson system). The nonrelativistic equations above were derived earlier in \cite{Adshead:2021kvl,Jain:2021pnk}. For a generalization to the spin-$s$ case, also see \cite{Jain:2021pnk}. In the future, it might also be interesting to systematically explore the relativistic corrections to this multicomponent system in the nonlinear regime \cite{Salehian:2021khb}.

\subsubsection{Conserved Quantities}
Note that in our convention, the number density, mass density, and spin density are 
\begin{align}
    \mathcal{N}(t,\bx)=\bPsi^\dagger\bPsi,\qquad\rho(t,\bx)=m\bPsi^\dagger\bPsi,\quad{\textrm{and}}\quad \bm{s}=i\hbar\bPsi\times\bPsi^\dagger.
\end{align}
The conserved quantities associated with our non-relativistic VDM are:
\begin{align}
\label{eq:conserved}
    N&=\int \mathrm d^3x \bPsi^\dagger\bPsi,\quad \textrm{and}\quad M=mN,\qquad\qquad\qquad\qquad\,\,\, \textrm{(particle number and rest mass)}\\
    E&= \int\mathrm{d}^3x\Bigl[\frac{\hbar^2}{2m}\,\nabla \bPsi^{\dagger}\cdot\nabla \bPsi- \frac{Gm^2}{2}\bPsi^{\dagger}\bPsi\int \frac{\mathrm{d}^3y}{4\pi|{\bm{x}}-{\bm{y}}|}\bPsi^{\dagger}({\bm{y}})\bPsi(\bm{y})\Bigr]\,,\qquad \qquad\quad\textrm{(energy)}\\
    \bS&=\hbar\int \mathrm d^3x\,i\bPsi\times\bPsi^\dagger\,,\quad\qquad\qquad\qquad\qquad\qquad\qquad\qquad\quad\, \textrm{(spin angular momentum)}\\
    \bm{L}&=\hbar\int \mathrm d^3x
    \,\Re\left(i\,\bPsi^{\dagger}\nabla\bPsi\times \bx\right).\,\,\qquad\qquad\qquad\qquad\qquad\qquad\textrm{(orbital angular momentum)}
\end{align}
Note that the spin and orbital angular momentum are separately conserved in the non-relativistic system. Importantly, by definition, spin angular momentum is identically zero for SDM (but not VDM). For details of the non-relativistic action and conserved quantities for a general spin-$s$ bosonic field (including VDM) see \cite{Jain:2021pnk}.

\subsubsection{Fluid equations}

We can also transform our multicomponent SP system eq.~\eqref{eq:SP_general} into a set of three, coupled fluid equations (following the Madelung transform commonly used in SDM \cite{1927ZPhy...40..322M}). With the following field re-definition, $\psi_j=\sqrt{\rho_j/m}\,e^{iS_j}$, and the velocity $\bm{u}_i \equiv \hbar\nabla S_i/m$, we have
\begin{align}
    \frac{\partial\rho_j}{\partial t} + \nabla\cdot(\rho_j \bm{u}_j) = 0\,,\qquad
    \frac{\partial\bm{u}_j}{\partial t} + (\bm{u}_j\cdot{\nabla})\bm{u}_j = \frac{1}{m}\nabla(Q_j - m\Phi),\quad\textrm{where}\quad j=1,2,3
\end{align}
where $Q_j = (\hbar^2/2m)\nabla^2\sqrt{\rho_j}/\sqrt{\rho_j}$. In terms of the Madelung variables, the spin density $s_i=i(\hbar/m^2)\epsilon_{ijk}\sqrt{\rho_j\rho_k}e^{i(S_j-S_k)}$. The vorticity for each of the three fluids $\bm{\omega}_j=\nabla\times \bm{u}_j=0$ if $\rho_j\ne0$. Note that zero vorticity does not imply zero spin density. If $\bm{\omega}_i\ne 0$ for some fixed $i$ (with ${\bm{\omega}}_{j\ne i}=0$), then $\bm{s}=s_i\hat{i}$. 

We numerically solve eq.~\eqref{eq:SP_general}, but the conservation/fluid equations can be useful in gaining physical intuition for the behaviour of the system (including for example, vortices \cite{Hui:2020hbq} in three  fluids.). 
\subsubsection{Scales and scalings}
The SP system (Eq.~\eqref{eq:SP_general}) has certain scaling symmetries, which significantly increases the generality of the results. Specifically, if we have a solution $\bPsi(t,\bx)$ for a system with total mass $M$ and vector boson mass $m$, then $\bPsi_{\beta\gamma}(t,\bx)=\beta^{5/2}\gamma^2\bPsi(\gamma^2\beta^3t,\gamma \beta^2\bx)$ is a solution for a system with mass $\gamma M$ and $\beta m$. Moreover,
\begin{equation}
\{m,M\}\rightarrow\{\beta m,\gamma M\}\Longrightarrow\{t,r,\rho,\bS\}\rightarrow\{ t/(\gamma^2\beta^3),r/(\gamma\beta^2),\gamma^4\beta^6\rho,(\gamma/\beta)\bS\}.
\end{equation}
For ease of comparison with astrophysical scales, let us define 
\begin{align}
m_{20}\equiv \frac{mc^2}{10^{-20}\,{\rm eV}},\quad \mathcal{M}_5\equiv \frac{M}{2.3\times 10^5M_\odot}\,.
\end{align}
The Compton length and time scales are then given by
\begin{align}
\lambda_m=\hbar/(mc)=6.4\times 10^{-7}\, \textrm{kpc}/m_{20},\quad\tau_m=\hbar/(mc^2)=2.1\times 10^{-3}\, \textrm{yr}/m_{20}.
\end{align}

\subsection{Wave Interference}
\label{sec:interference}
Consider the density resulting from the superposition of two unit amplitude plane waves in a spin-$s$ field ($s=0$ for SDM and $s=1$ for VDM): $\bPsi_a(\bx)=V^{-1/2}\,\bepsilon^{(s)}_{a}e^{i\bk_a\cdot\bx}$, where $V$ is the volume, $a=1,2$, and $\bepsilon^{(s)}_{a}$ is a unit complex vector:
\begin{equation}
  |\bPsi_a(\bx)+\bPsi_b(\bx)|^2=2\,V^{-1}\left(1+\Re\left[\bepsilon^{(s)\dagger}_{a}\cdot\bepsilon^{(s)}_{a}e^{-i(\bk_a-\bk_b)\cdot\bx}\right]\right)=2\,V^{-1}\left(1+\inter_{(s)}
  \right).  
\end{equation}
Here, the factor of $2$ is the number of waves and ${\rm int}_{(s)}$ is the interference term. Without loss of generality, we set $\bx=0$. The interference term is simply the cosine of the angle between the two waves (unit vectors in $2$ dimensions for SDM and $6$ for VDM), with their heads lying on a unit $4s+1$ sphere. Assuming a uniform distribution on the sphere, the cosine of the angle between these waves $\inter_{(s)}\equiv\eta=\cos\theta$ has the following distribution $p^{(s)}(\eta)=\pi^{-1/2}\left\{\Gamma(2s+1)/\Gamma(2s+1/2)\right\}(1-x^2)^{(4s-1)/2}$. While the mean is zero, the standard deviation 
\begin{equation}
    \sqrt{\langle \inter_{(s)}^2\rangle}=\frac{1}{\sqrt{2(2s+1)}},\qquad \textrm{with}\qquad \frac{\sqrt{\langle \inter_{(1)}^2\rangle}}{\sqrt{\langle \inter_{(0)}^2\rangle}}=\frac{1}{\sqrt{3}}.
\end{equation}
That is, interference decreases for higher spin fields. This is a reflection of the intuitive fact that in a larger component field, orthogonal components do not interfere. This simple fact has important implications for differences between VDM and SDM.

\subsection{Solitons}
Ground state solitons in VDM are characterized by the ``chemical potential" $\mu$ and a unit complex 3-vector $\bepsilon$ (with $\bepsilon^\dagger \bepsilon=1$):
\begin{equation}
\label{eq:VDMSoliton}
    {\bPsi}_{\rm sol}({t},{\bx})={\psi}_{\rm sol}({\mu},{r})e^{i{\mu c^2}{t}/\hbar}\bepsilon,
\end{equation}
where ${\psi}_{\rm sol}$ is a real valued spherically symmetric function that satisfies
\begin{equation}
\label{eq:MS_mu}
    -{\mu c^2}{\psi}_{\rm sol}=-\hbar^2\frac{{\nabla}^2}{2m}{\psi}_{\rm sol}+m{\Phi} {\psi}_{\rm sol},\quad {\nabla}^2{\Phi}=4\pi Gm{\psi}_{\rm sol}^2.
\end{equation}
Note that the profile for a VDM soliton satisfies the same time-independent equation as a SDM soliton. The mass and spin of this soliton are given by 
\begin{equation}
{M}_{\rm sol}\approx 60.7\frac{\mpl^2}{m}\sqrt{\frac{\mu}{m}},\quad{\bS}_{\rm sol}\approx i(\bepsilon\times\bepsilon^\dagger)60.7\frac{\mpl^2}{m^2}\sqrt{\frac{\mu}{m}}\hbar.
\end{equation}

The special cases of maximally polarized solitons configurations are given by $\bepsilon_{(0)}=\hat{z}$  and $\bepsilon_{(1)}=(\hat{x}+ i\hat{y})/\sqrt{2},$  along with their spatial rotations \cite{Jain:2021pnk}. Configurations with $\bepsilon=\bepsilon_{(0)}$ are linearly polarized, with zero total spin angular momentum \cite{Adshead:2021kvl}. Whereas, configurations with $\bepsilon=\bepsilon_{(1)}$ have a maximal spin angular momentum $|{\bS}_{\rm sol}|=\hbar{M}_{\rm sol}/m$. For all other solitons, we expect the spin angular momentum to lie between these maximal values. That is $0\le |{\bS}_{\rm sol}|\le \hbar{M}_{\rm sol}/m$ \cite{Jain:2021pnk}. Note that these solitons (with and without spin) have lower energies than those with spherically symmetric vector fields vanishing at the origin~\cite{Brito:2015pxa,DiGiovanni:2018bvo}, at least in the nonrelativistic limit.

In \cite{Schive:2014dra}, the scalar soliton profile was parameterized by a characteristic width $r_c$, so that the density and mass can be characterized as
\begin{equation}
\label{eq:core_fit}
    {\rho}_{\rm sol}(r)\approx 1.9\times 10^7 m_{22}^{-2}\frac{\left({\rm kpc}/{r}_c\right)^4}{(1+0.091(r/{r}_c)^2)^8}\frac{M_\odot}{\kpc^3},\qquad M_{\rm sol}\approx 2.2\times 10^8 \left(\frac{\kpc}{r_c}\right)m_{22}^{-2}M_\odot.
\end{equation}
Using eq.~\eqref{eq:core_fit} and eq.~\eqref{eq:MS_mu}, we have $r_c=6.8\times 10^{-5}m_{22}^{-1}\sqrt{m/\mu}\,\kpc$. 
For the solitons in our simulations we typically have ${\mu/m}\sim 10^{-12}$.

We re-iterate that the soliton profile is characterized by the same function in VDM and SDM. The analysis is identical with  $\bepsilon\rightarrow e^{i\varphi}$ (a phase) for SDM. Also note that the  vector solitons discussed above are gravitationally bound; it is also possible to have vector solitons (vector oscillons \cite{Zhang:2021xxa}) which are bound by attractive self-interactions. In such self-interaction supported solitons, linearly polarized solitons have a lower energy compared to the circularly polarized ones (at fixed particle number). As in the case of scalars (see for example, \cite{Chavanis_2011,Chavanis:2011zm}), it will be possible to construct non-relativistic solitons with a combination of self-interactions and gravity \cite{Jain:2022nqu}.

\section{Two soliton mergers}\label{sec:two_merger}
In this section we explore the merger of 2 solitons, as a warm up to the $N$-soliton case. For simplicity, we restrict ourselves to head-on collisions only. 

We begin with two identical VDM solitons with a characteristic radius $r_c\approx 1\,{\rm kpc}\times (\mathcal{M}_4 m_{20}^2)^{-1}$ and separation of $\approx 10{r}_c$, with masses $M_{\rm sol,1}=M_{\rm sol,2}$. We define $\mathcal{M}_4=(M/2.2\times 10^4 M_\odot)$. We give each of them a small $v/c\approx 3.3\times 10^{-7}\,\mathcal{M}_4m_{20}$ velocity towards each other (the typical velocity expected in our $N$ soliton simulations at this distance). Each VDM soliton has its own complex unit vector $\bepsilon_{1,2}$ (see eq.~\eqref{eq:VDMSoliton}). After the collision, if a new soliton forms, it has a mass 
$M_{\rm sol,f}=f_{\rm v}(M_{\rm sol,1}+M_{\rm sol,2})$, 
where $1-f_{\rm v}$ is mass fraction that does not end up in the final soliton. Here, we are partly repeating similar analyses carried out for solitons in SDM \cite{Du:2016zcv,Hertzberg:2020dbk}. We note that in our analysis, the merged configuration is often highly excited, and it is not always clear that it is a settled soliton within the duration of the simulation. Nevertheless, we will continue to refer to the merged object as a soliton for simplicity in this section.\footnote{Since the merged object is not spherically symmetric and can have large excitations, we average the spatial profile of a coalesced state for many instances in time to get a time averaged spatial profile. This profile is then fit to a soliton profile to define the width $r_c$. The merged soliton mass is calculated from this $r_c$.  The ratio of this final soliton mass to the total initial mass is what we define as $f$. In the context of points 1. and 2. discussed below, we found $r_c\approx\{.824,.812,.826,.816,.806,.812,.827\}\,(\mathcal{M}_4m_{20}^{2})^{-1}\rm kpc$ for the scalar case, and  $r_c\approx\{0.89,0.95,0.9,0.91,0.87,0.81,0.92,0.81,0.89,0.83,9.92,0.93\}\,(\mathcal{M}_4m_{20}^{2})^{-1}\rm kpc$ for the vector case based on 7 and 12 separate simulations respectively. For these 2 soliton mergers, we start them at a separation of $10 (\mathcal{M}_4m_{20}^{2})^{-1}\rm kpc$, and give them a small boost to account for the speed that such solitons would have in our $N$ soliton merger simulations at this separation.}

Along with the mass loss fraction, we also investigate the evolution of intrinsic spin for the VDM case. As with mass, while the total spin angular momentum is conserved, the final soliton need not carry all of the initial spin angular momentum. This naturally generates the possibility of creating a merged object with intrinsic spin angular momentum, even if the two initial solitons had none.

We carried out about $\sim 10$ runs for the case with $\bepsilon_1=\bepsilon_2 e^{i\varphi}$, and $10$ runs for $\bepsilon_1\ne \bepsilon_2e^{i\varphi}$. Our simulations have periodic boundary conditions, so instead of calculating the mass that escapes, we focus on the merged object. The merged configuration typically shows large oscillations, which we average over and fit with a single parameter profile \eqref{eq:core_fit}, to obtain a value for $r_c$ and $M_{\rm sol,f}$. We also kept track of the intrinsic spin angular momentum, which is calculated using eq.~\eqref{eq:conserved} in the region of interest. Our key findings are as follows:
\begin{enumerate}
\item If $\bepsilon_1=\bepsilon_2 e^{i\varphi}$, the collision is identical to the collision of two solitons in SDM with a  phase difference $\varphi$ between them (confirming the expectation in \cite{Jain:2021pnk}). In this case the solitons merge to form another soliton $0\le\varphi<\pi$ with $f_{\rm v}=f_{\rm s}\approx 0.61\pm0.01$. This is based on the inferred $r_c$ of the final object after angular and temporal averaging.
\item If $\bepsilon_1\ne \bepsilon_2 e^{i\varphi}$, then in general, such a collision cannot be mimicked by binary soliton collisions in SDM \cite{Jain:2021pnk}. Restricting ourselves to the cases where $\bepsilon^\dagger_1\cdot\bepsilon_2>0$ (so that the solitons can merge relatively quickly due to constructive interference), we find that $f_{\rm v}\approx 0.56\pm 0.03$.
\item To investigate the evolution of spin angular momentum, we consider the case where $\bepsilon_1=\hat{x}$ and $\bepsilon_2=i\hat{y}$. In this case, both initial solitons have zero spin angular momentum (they are linearly polarized). At the end of the merger, we find that the final core  has a spin angular momentum $\bS_{\rm c,f}\propto \hbar\hat{z}$. The field which is not in the core carries an equal amount of (opposite signed) spin angular momentum.  We note that the core is not quite a soliton for this particular case; density of different components are not spherically symmetric around a common origin.\footnote{If instead of $\bepsilon_2=i\hat{y}$ we had chosen $\bepsilon_2=\hat{y}$, the dynamics of the fields would remain identical. However, the resulting spin density would not. More generally, if we have a solution $\bPsi$ of our system, then $U\bPsi$ is also a solution if $U^\dagger U=1$. Moreover the number density $\mathcal{N}$ remains the same. However, the spin density $\bm{s}$ does not remain the same in the general case. How does the spin density change? For $U=\mathcal{O}$, with $\mathcal{O}^T\mathcal{O}=1$, we have $\bm{s}\rightarrow \mathcal{O}\bm{s}$ as expected. However for  $U_{ij}\propto\delta_{ij}$, with $U_{jj}=ie^{\alpha_j}$, we have
$i\bPsi\times \bPsi^\dagger\rightarrow 2\Im[\epsilon_{ijk}e^{i\alpha_{ij}}\psi_i\psi_j^*]\ne U\bm{s}$ where $\alpha_{ij}\equiv\alpha_i-\alpha_j$.} 
\end{enumerate}
This is a preliminary investigation of 2 soliton mergers in VDM, and the dependence on impact parameters, angular momentum, initial spin, energy etc. needs to be investigated further.  We note that our $f_{\rm s}\approx 0.61$ is less than $\approx 0.7$ quoted in the literature \cite{Du:2016zcv}, which could be due to different ways in which the mass loss fraction is calculated as well as the initial relative velocities used. 

\begin{figure}[t!]
    \centering
      \includegraphics[width=6.5in]{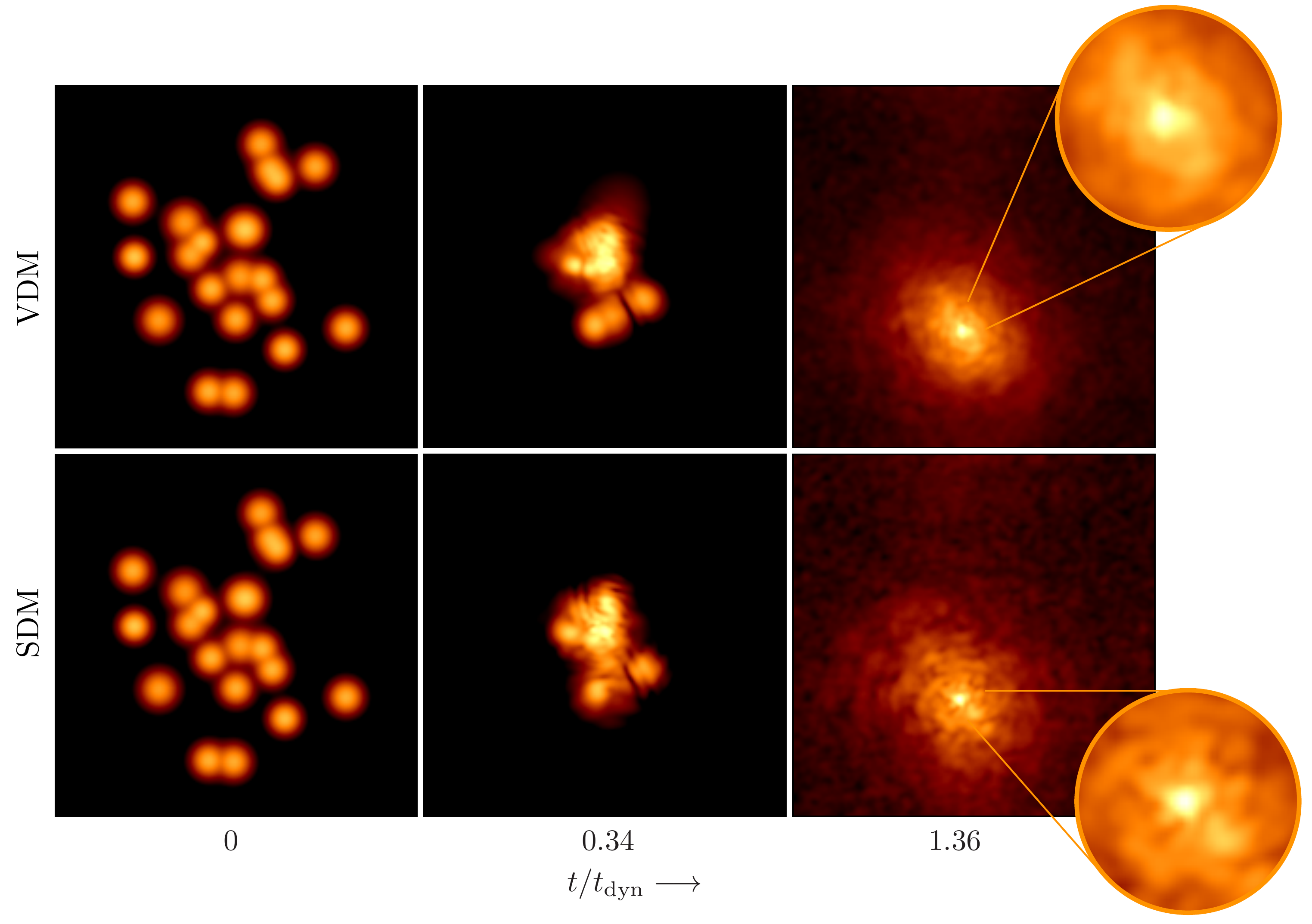}
\caption{\small{Starting with a collection of idealized halos (solitons), we eventually evolve to an approximately spherically symmetric configuration with a central core surrounded by a halo. Top row is vector dark matter (VDM), bottom row is scalar dark matter (SDM). For identical initial conditions in density, the VDM final central core is less dense and its halo shows less interference, as compared to SDM. The core to halo transition is also smoother in VDM compared to SDM. In the above images, the color represents the projected mass density in the simulation volume. Lighter colors correspond to higher mass densities.}}
      \label{fig:snapCompare}
\end{figure}
\section{Many soliton mergers}\label{sec:many_merger}
We begin with $N\sim \mathcal{O}[10]$ solitons whose positions are chosen randomly within our simulation volume. As we let the system evolve, gravitational interactions bring the solitons closer. Field interference and nonlinear evolution leads to a complex transient phase, after which, the density settles into an approximately spherically symmetric density configuration. The typical time-scale of this transient phase is less than the dynamical time scale $t_{\rm dyn}=1/\sqrt{G\bar{\rho}}$ of our systen. When comparing SDM and VDM, the initial density at each point in the simulation volume is always identical. Snapshots of the time evolution of SDM and VDM are shown in Fig.~\ref{fig:snapCompare}.

We consider the case where all the solitons have the same initial radii, as well as the case where we draw the radii from a Gaussian distribution. We also consider the case where we change the total mass $M_{\rm tot}$ while fixing the number of solitons, as well as the case where we fix the number of solitons and change the total mass. For SDM, the initial phase for each soliton is drawn from a uniform distribution between $0$ and $2\pi$. For VDM solitons, a complex unit vector $\bepsilon$ is a $6$ dimensional unit vector with its head uniformly distribution on a unit 5-sphere. 

\begin{figure}[t!]
    \centering
    \includegraphics[scale=.6]{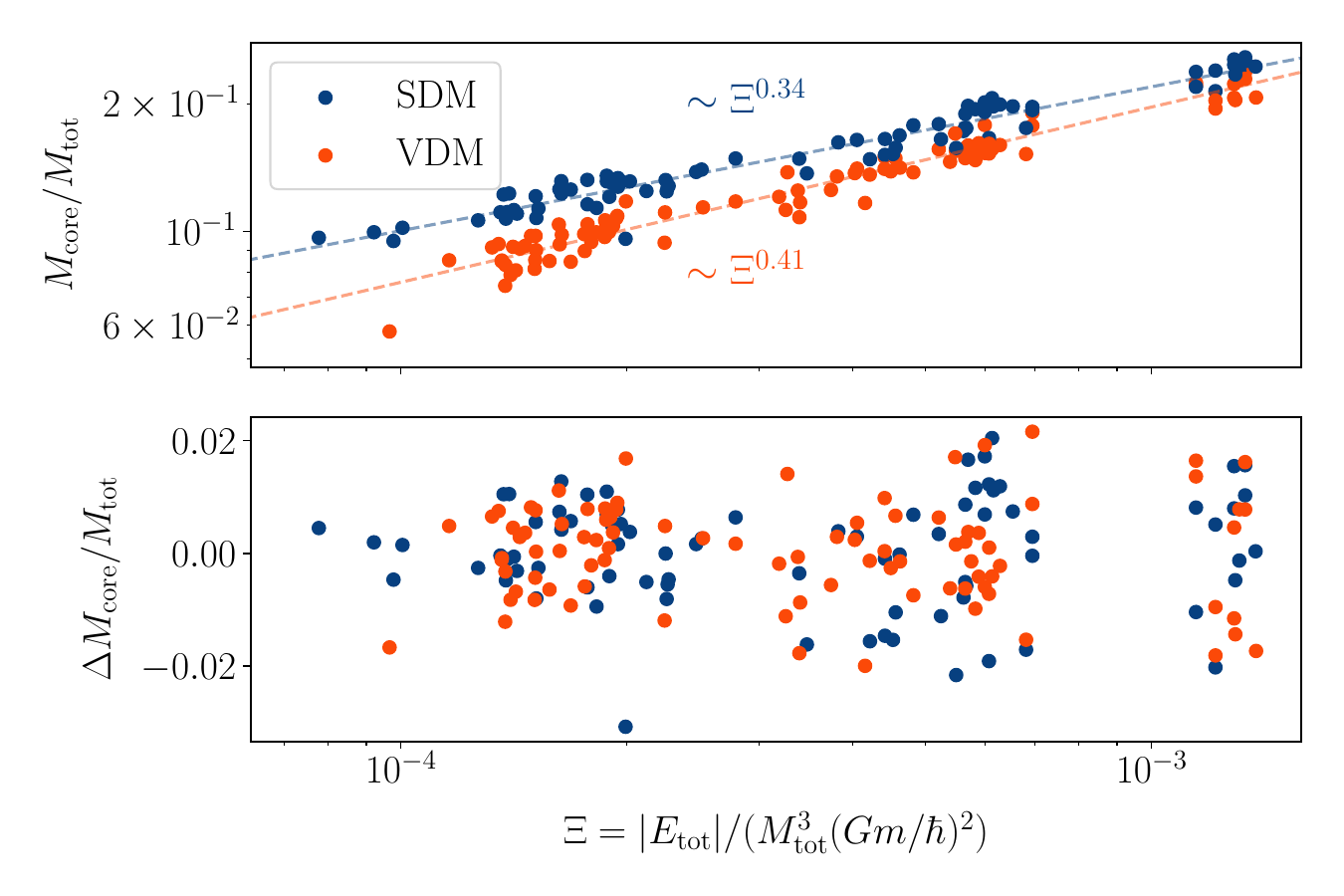}
    \caption{\small{The ratio of the final core mass to the total mass is plotted against the invariant quantity $\Xi$ for $\sim 80$ SDM and $\sim 80$ VDM configurations. The numerical power law fits, $M_{\rm core}/M_{\rm tot}=a\, \Xi^\alpha$, are also shown in the above panel. The fitted exponent $\alpha$ (with errors $\delta\alpha\lesssim 0.01$) obtained from the numerics are in agreement with expectations of simple analytic arguments related to repeated binary mergers. The bottom plot shows the deviation of $M_{\rm core}/M_{\rm tot}$ from the best fit lines. The average scatter, standard deviation of $\Delta  M_{\rm core}/M_{\rm tot}$ divided by the fit, in VDM is $\sim 15\%$ larger than SDM. 
    }}
    \label{fig:CoreByXi}
\end{figure}

\begin{figure}[t!]
    \centering
    \subfloat[]{
      \includegraphics[scale = .5]{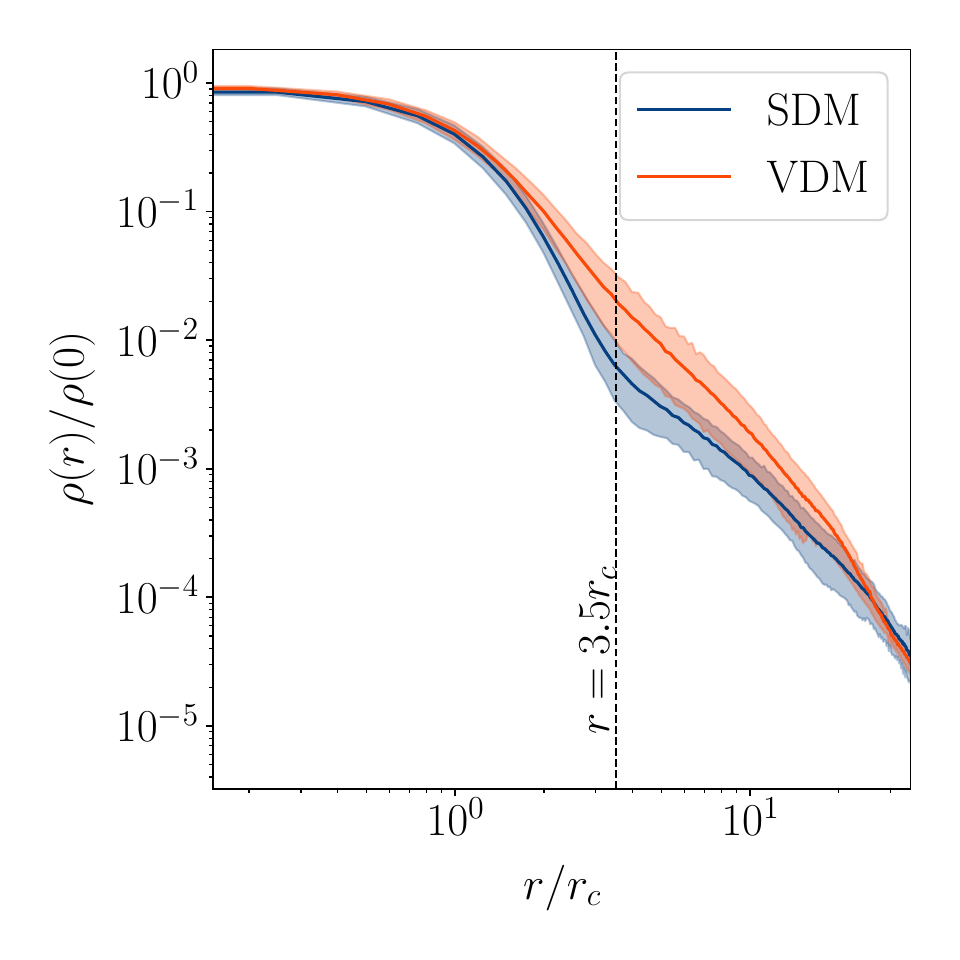}
      \label{fig:radialDensityAll}
    }
    \subfloat[]{
        \includegraphics[scale = .5]{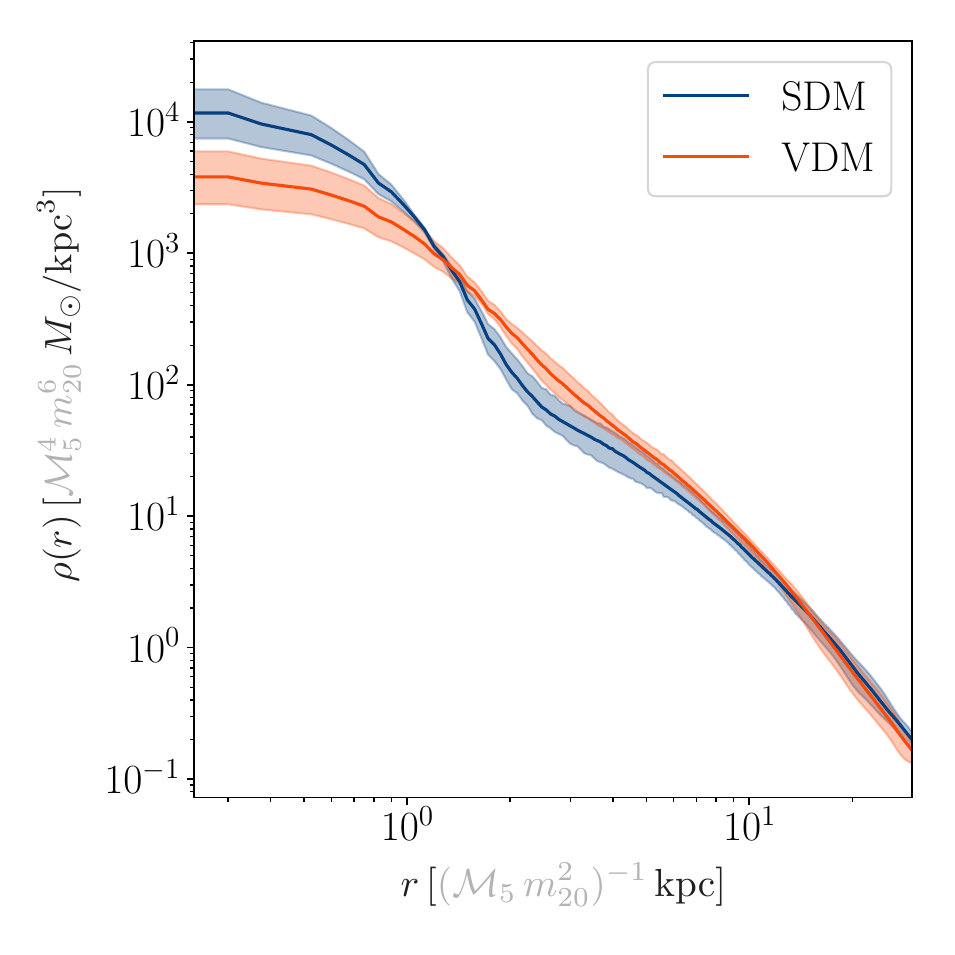}
      \label{fig:plotRD21_averaged}
    }
\caption{\small{Left panel \eqref{fig:radialDensityAll}: Angle-averaged, late-time central core+halo profiles for $\sim 80+80$ simulations spanning a range of initial conditions including different total mass, initial number of solitons, locations of solitons, phases and spins of solitons (i.e. $\Xi$ spans an order of magnitude). The radial coordinate and density are normalized by $r_c$ and $\rho(r=0)$ to highlight the differences in profile shape of VDM and SDM coalesced cores independent of the initial conditions. Solid lines indicate average over different simulations, the shaded regions indicate the spread in all profiles. A marker at $r/r_c \approx 3.5$ shows a general transition between core/halo regions in both SDM and VDM scenarios. Right panel \eqref{fig:plotRD21_averaged}: Final radial density from 10 simulations (time averaged over roughly 1 period of radial oscillations of the core), where the initial mass is narrowly distributed around $M_{\rm tot}=2.3\times 10^5 \; M_\odot\times \mathcal{M}_5$, the size of the simulation volume is $L=100\,{\rm{kpc}}\times (\mathcal{M}_5m_{20}^2)^{-1}$ and the number of initial solitons was fixed at 21. }}
\label{fig:radialDensity}
\end{figure}

\subsection{Core-halo mass}
In our simulations, we find a tight correlation between the mass of the final core and the total mass of the system: $M_{\rm core}/M_{\rm tot}\propto \Xi^{\alpha}$, where $\Xi$ is a measure of the total energy of the system, and $\alpha$ is different for VDM and SDM. This correlation is shown in Fig.~\ref{fig:CoreByXi}. Below, we present an explanation for the observed relationship.\footnote{Note that from simulations, $M_{\rm core}$ is obtained by first fitting the central density region with \eqref{eq:core_fit} to obtain an $r_c$, and then including mass withing a sphere of radius $2r_c$. For a soliton, $M_{\rm core}=M(r<2r_c)=(3/4)M_{\rm sol}$. Also note that $r_c$ for the core is not fixed in time, and towards the end of the simulations shows oscillatory time variations of roughly $20\%$.}

Beginning with $N$ solitons of mass $M_{\rm sol}^{\rm i}$ each, and distributed randomly throughout the box, the total energy is (scaled to yield a dimensionless scale-invariant measure $\Xi$)
\begin{align}
\label{eq:energytot}
    \Xi\equiv \frac{|E_{\rm tot}|}{M_{\rm tot}^3(Gm/\hbar)^2}  &\approx \frac{1}{M_{\rm tot}^3(Gm/\hbar)^2}\left[N\frac{G(M_{\rm sol}^{\rm i})^2}{2R_{\rm sol}^{\rm i}} + (1.88)N(N-1)\frac{G(M_{\rm sol}^{\rm i})^2}{L}\right],\\
    &\approx \frac{1}{20N^2}.
\end{align}
In the first line, $L$ is the box size and $R_{\rm sol}^{\rm i} \ll L$ is the initial solitons' radius. In the last equality, we have assumed that the first term in eq.~\eqref{eq:energytot} dominates over the second.\footnote{Note that $R_{\rm sol}^{\rm i}\equiv 9.95 \hbar^2/(GM_{\rm sol}^{\rm i} m^2)$ contains $99\%$ of the soliton's mass, and we also include gradient contributions to the individual soliton's energy. The factor of $1.88$ in the second term arises from an ensemble average over initial locations of the solitons. We have ignored corrections due to overlap of solitons. For our system, the second term is only marginally smaller than the first.}

To relate the final soliton mass to the initial one, we follow a simplified version of the arguments used in \cite{Du:2016zcv}. See our Appendix~\ref{sec:EstSolMass} for details. Suppose that for $N$ initial solitons, $N_{\rm major}\equiv \gamma N$ are major binary mergers, and in each major merger a fraction $f$ of the progenitors mass is contained in the resulting merged soliton. In this case, the final soliton mass after $\gamma N$ major mergers is given by $M_{\rm sol}^{\rm f}\approx(\gamma N)^{\log_2(2f)}{M_{\rm sol}^{\rm i}}=(\gamma N)^{\log_2(f)}M_{\rm tot}$. Solving for $N$ and using this result in eq.~\eqref{eq:energytot}  we arrive at 
\begin{align}
\label{eq:Mc/M}
    \frac{M_{\rm core}}{M_{\rm tot}}\approx a\, \Xi^{-\log_2(\sqrt{f})}\qquad\textrm{where}\qquad a=\frac{3}{4}\frac{\gamma}{f}\left(\frac{\gamma}{\sqrt{5}}\right)^{\log_2(f)}\,.
\end{align}
Here, we used $M_{\rm core}=M_{\rm sol}(r<2r_c)=(3/4)M_{\rm sol}^{\rm f}$. The fraction $f$ depends on whether we have VDM or SDM.
From the previous section on two-soliton collisions we found that for VDM, $f_{\rm v}\approx 0.56\pm0.03$, and for SDM, $f_{\rm s}\approx 0.61\pm0.01$. These $f$s translate to the following exponents of $\Xi$ for VDM and SDM:
\begin{align}-\log_2(\sqrt{f_{\rm v}})\approx 0.42\pm0.04\qquad-\log_2(\sqrt{f_{\rm s}})\approx 0.36\pm 0.02,
\end{align} which match the numerical obtained exponents  well (see Fig.~\ref{fig:CoreByXi}). We have not calculated the fraction of major mergers $\gamma$, so we do not try to estimate the coefficient $a$. Nevertheless, if we assume that $\gamma$ is the same for SDM and VDM (with $1/4\lesssim\gamma\lesssim 1$), then the analytic estimate captures the result that $M_{\rm core}/M_{\rm tot}$ for VDM should be lower than SDM for a given $\Xi$ (in the range shown in the figure), and that the exponent of $\Xi$ for VDM should be higher than that for SDM. 

We expect that the tight correlation between core and halo mass in VDM will be disrupted for cosmological simulations (as opposed to our idealized halo mergers), as was the case for SDM \cite{May:2021wwp,Chan:2021bja}.

\subsection{Late-time density profiles}
We will focus on the  (angular averaged) radial density profile of the late time configuration in our simulation volume. We find that the central region is well described by a soliton-like core, which eventually transitions into a power law tail at larger radii. Starting with identical initial density distributions in VDM and SDM, we typically find that the central solitonic core is less dense for VDM.

Fitting for the cores with a soliton profile (see \eqref{eq:core_fit}), we extract a core radius $r_c$ for each run. For a soliton, this $r_c$ uniquely fixes the central density $\rho_c\propto r_c^{-4}$. For ease of comparison between simulations with different initial conditions (which include different (total) initial masses, number of solitons, distribution of radii, locations of solitons etc.) we scale the density and radius of the final configuration by $\rho_c$ and $r_c$. The resulting collection of density profiles are shown in Fig.~\ref{fig:radialDensityAll}. 
\\ \\
\noindent The key observations common to VDM and SDM based on these normalized density profiles are as follows:
\begin{itemize}
\item For both VDM and SDM, a soliton-like core is clearly visible for $r/r_c\lesssim 1$. At $r/r_c\gtrsim 1$, the profile starts dropping rapidly. A transition from core to an $r^{-3}$ tail, occurs between $1\lesssim r/r_c\lesssim 10$. For $r/r_c\gtrsim 10$, we see $\rho/\rho_c\propto (r/r_c)^{-3}$.\footnote{The periodic box makes it difficult to trust the detailed power law when the radii become comparable to the size of the box. The $r^{-3}$ is not robust at radii comparable to the size of the simulation volume.} The transition region is qualitatively delineated by $r/r_c=3.5$ in Fig.~\ref{fig:radialDensityAll}.
\end{itemize}
The key distinguishing feature between VDM and SDM is that
\begin{itemize}
   \item the transition from the soliton-like profile to $r^{-3}$ profile occurs a lot more smoothly in VDM compared to SDM. This shape information is relatively independent of our initial conditions.
\end{itemize}
The power law regime joins the soliton profile for $r/r_c\sim 1$ in case of VDM. For SDM, the soliton-like profile persists for $r/r_c \gtrsim 1$, after which there is a transitory power law (shallower than $r^{-3}$), before joining the $r^{-3}$ tail at $r/r_c\sim 10$.

\begin{figure}[t!]
    \centering
    \subfloat[]{
      \includegraphics[scale = .5]{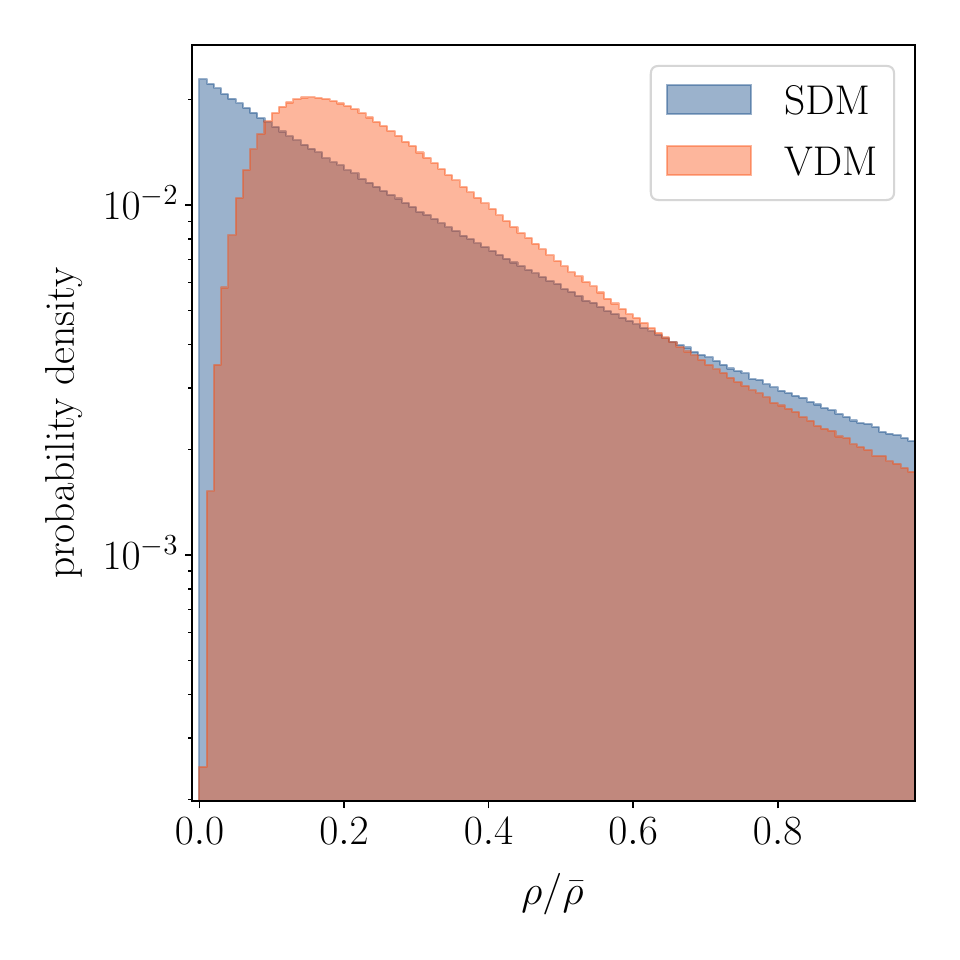}
      \label{fig:rawDensity}
    }
    \subfloat[]{
      \includegraphics[scale = .5]{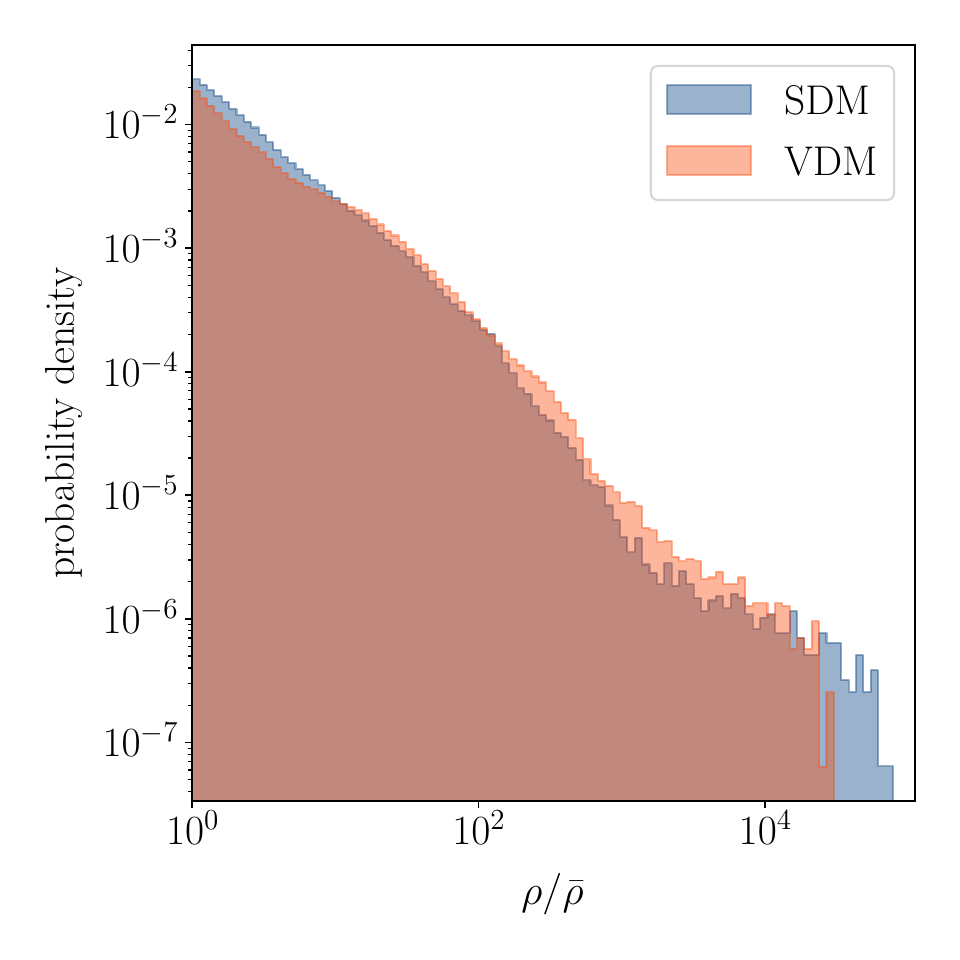}
      \label{fig:rawDensityFT}
    }
\caption{\small{The normalized histogram of density (probability density function) from a SDM and a VDM simulation, each with ~$M_{\rm tot}=2.2\times10^5 \; M_\odot\times \mathcal{M}_5$.  At low densities compared to the mean ($\rho/\bar{\rho}\ll1$), we see important qualitative differences between SDM and VDM \eqref{fig:rawDensity} (left panel), with a dearth of ultra-low density regions in VDM due to reduced interference. In the right panel, \eqref{fig:rawDensityFT}, we see a lack of ultra-high densities in VDM, again due to reduced interference. The general PDF shapes of SDM and VDM densities, especially at low densities, are found to be robust for a wide range of initial conditions.}}
\label{fig:rawDensities}
\end{figure}
\subsection{Density PDF}
As discussed earlier, interference effects are reduced in VDM compared to SDM. As a result, we expect fewer deviations from the average density at low and very high densities for VDM compared to SDM. This is seen in the PDF of density extracted from our simulations (see Fig.~\ref{fig:rawDensity} and Fig.~\ref{fig:rawDensityFT}). The horizontal axis is normalized by the average density in the box.

To gain further (still qualitative) understanding of the low-density shape difference (near $\rho/\bar{\rho}\rightarrow0$) between VDM and SDM, we consider the density resulting from the superposition of a large number of unit amplitude waves with random phases. In that case the density pdf for a spin-$s$ field is a Gamma distribution $\mathcal{P}^{(s)}(\rho/\bar{\rho})\sim {(2s+1)^{(2s+1)}}/{\Gamma[2s+1]}\left({\rho}/{\bar{\rho}}\right)^{2s}e^{-(2s+1)\rho/\bar{\rho}}$. 
In particular, for $s=0$ (SDM) and $s=1$(VDM) and for $\rho/\bar{\rho}\ll 1$, we have
\begin{equation}
\mathcal{P}^{(0)}(\rho/\bar{\rho})\sim 1-\frac{\rho}{\bar{\rho}}+\hdots,\qquad \mathcal{P}^{(1)}(\rho/\bar{\rho})\sim \left(\frac{\rho}{\bar{\rho}}\right)^{\!2}+\hdots
\end{equation}
which explains the qualitative behavior of the pdf at low densities in VDM and SDM seen Fig.~\ref{fig:rawDensity}.  We caution that assuming unit amplitudes for all waves is not justified, and we do not expect the numerical pdf to agree with our analysis above quantitatively. Nevertheless, this qualitative understanding and shape will already be useful for observational implications (see Section~\ref{sec:end}).

At high densities (Fig.~\ref{fig:rawDensityFT}), we see that VDM has a shorter tail compared to SDM which is again a direct consequence of reduced interference in VDM. The extreme high density regions are found in the central core for both VDM and SDM.
\begin{figure}[t!]
    \centering
    \subfloat[]{
      \includegraphics[scale = .48]{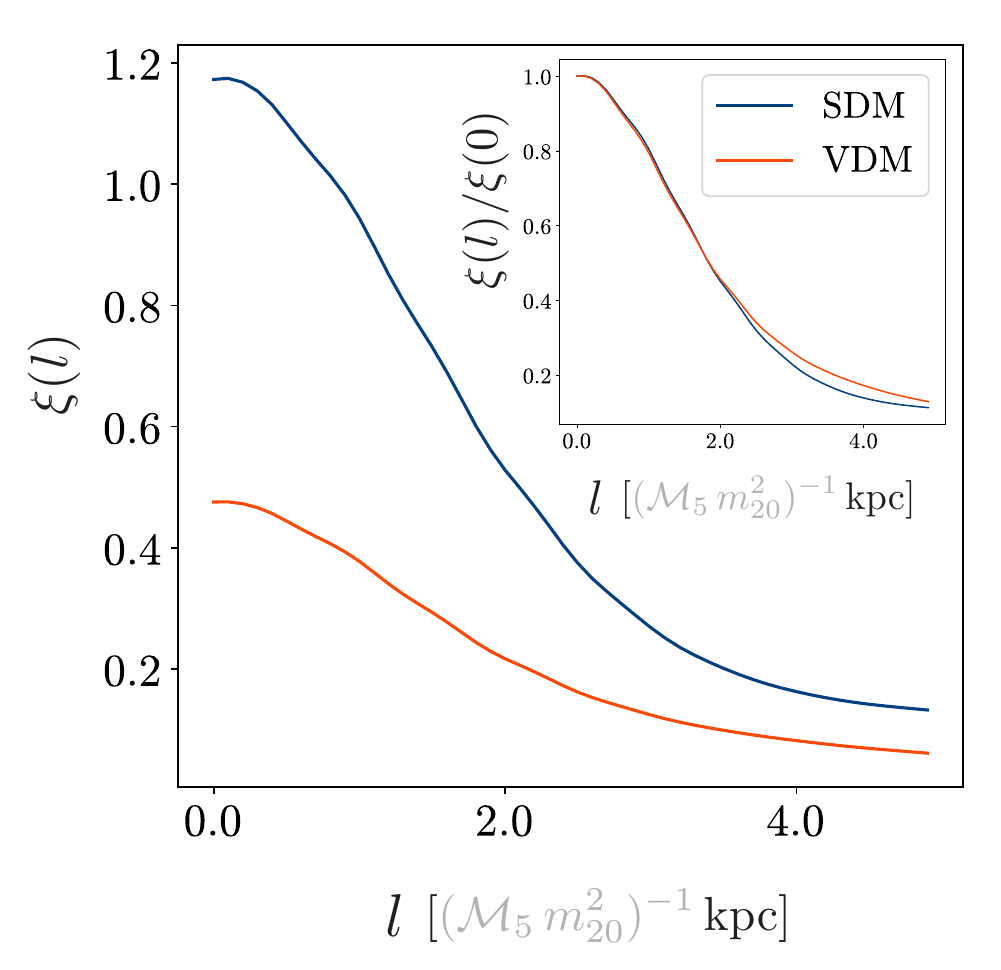}
      \label{fig:Correlation}
    }
    \subfloat[]{
      \includegraphics[scale = .49]{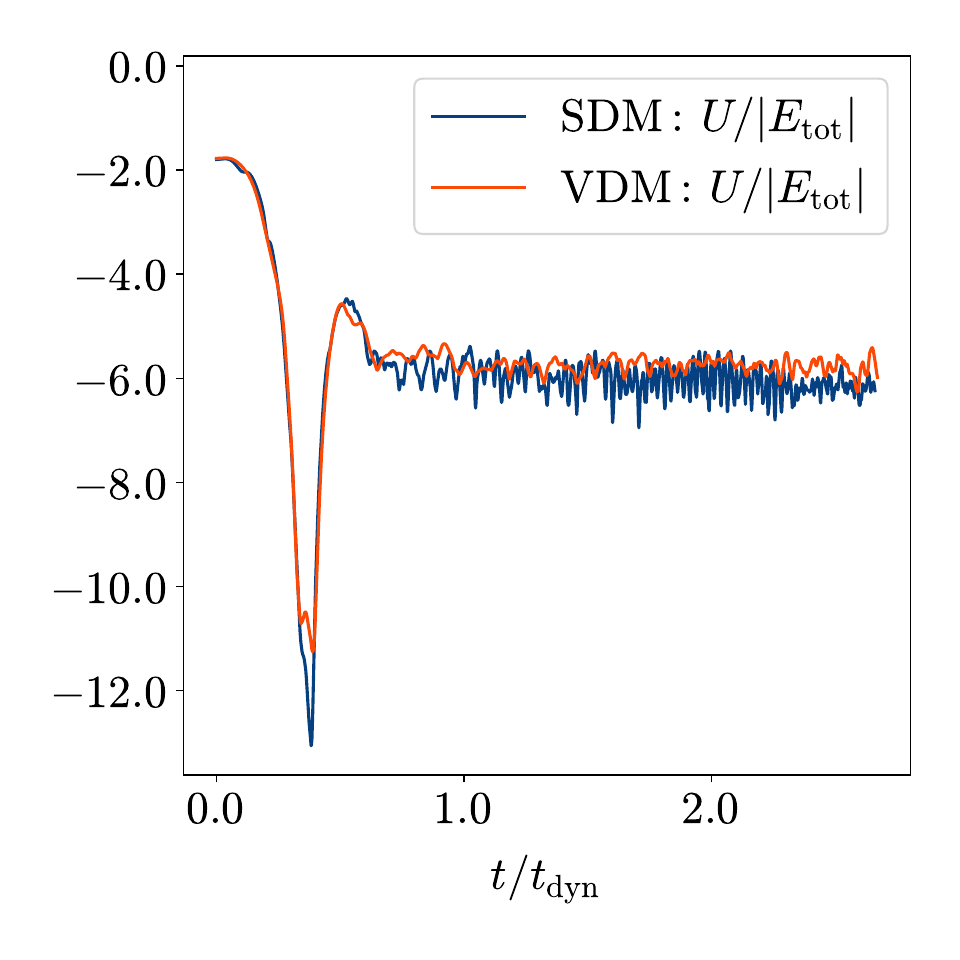}
      \label{fig:ES}
    }
    \caption{\small{In Fig.~\ref{fig:Correlation}, we show the two point correlation function of fractional density fluctuations in the halo (c.f.~\eqref{eq:corr_volavg}) for a single simulation with $M_{\rm tot}\approx 2.2\times 10^5\,M_\odot \times \mathcal{M}_5$ and $N=21$ at late times. The central core is removed before calculating the correlation function. The lower amplitude of the VDM curve reflects the smaller variance of the density fluctuations, whereas the similarity in width of the correlation functions (inset) shows that the physical size of the interference granules is similar in SDM and VDM. In Fig.~\ref{fig:ES},  we show the potential energy in the system as a function of time. The potential energy (and kinetic energy) in SDM shows more temporal fluctuations compared to VDM. The kinetic energy can be obtained from $K/|E_\mathrm{tot}|=1-U/|E_{\rm tot}|$. }}
    \label{fig:2ptE}
\end{figure}

\subsection{Two point correlation functions}
To understand the typical amplitude and characteristic scale of density fluctuations in the halo surrounding the central core, we construct the two point correlation function
\begin{align}\label{eq:corr_volavg}
    \xi(\bm{l})=V^{-1}\int_V d^3r\,\delta(\bm{r})\delta(\bm{r}+\bm{l})\,,
\end{align}
where we have removed the central high density core to reduce the influence of the central core, so that $V$ is the volume enclosed between $r \sim 8r_c$ and $r \sim 20r_c$ from the center of the core. Here, $\delta(\bm{r})= (\delta\rho/\rho)(\bm{r})\equiv [\rho(\bm{r})-\bar{\rho}(r)]/\bar{\rho}(r)$ where $\bar{\rho}(r)$ is the angular averaged density. Fig.~\ref{fig:Correlation} shows the angular averaged $\xi(\bm{l})$.\footnote{This figure has changed from our previous version where we used the total density rather than the fractional deviation of the density in the correlation function.} The VDM correlation function at $l=0$ is a measure of the variance of $\delta(\bm{r})$. We find $[\xi(0)]_{\rm vdm}/[\xi(0)]_{\rm sdm}\sim 1/3$ in agreement with our analysis in Section \ref{sec:interference}, indicating a reduced interference in VDM compared to SDM. When the correlation functions in SDM and VDM are scaled by the $[\xi(0)]_{\rm vdm}$ and $[\xi(0)]_{\rm sdm}$ respectively, we find that they overlap, and have the same width. 
We take this to be an indication that the spatial size of the interference granules in SDM and VDM are the same in the outer region of the halo, $r\gtrsim 10r_c$, where the mass enclosed as a function of radius becomes identical in SDM and VDM.\footnote{Using the radial profile of mass density, we find that the circular velocity, and hence approximate de Broglie scale in the outer regions of the halo is $\lambda_{\rm dB}\sim (\mathcal{M}_5^2m_{20}^2)^{-1}1{\rm \,kpc}$.}

Another measure of the density structure in VDM and SDM is the power spectrum, which is provided in \ref{app:PS}.
\begin{figure}[t!]
    \centering
    \includegraphics[scale=.3]{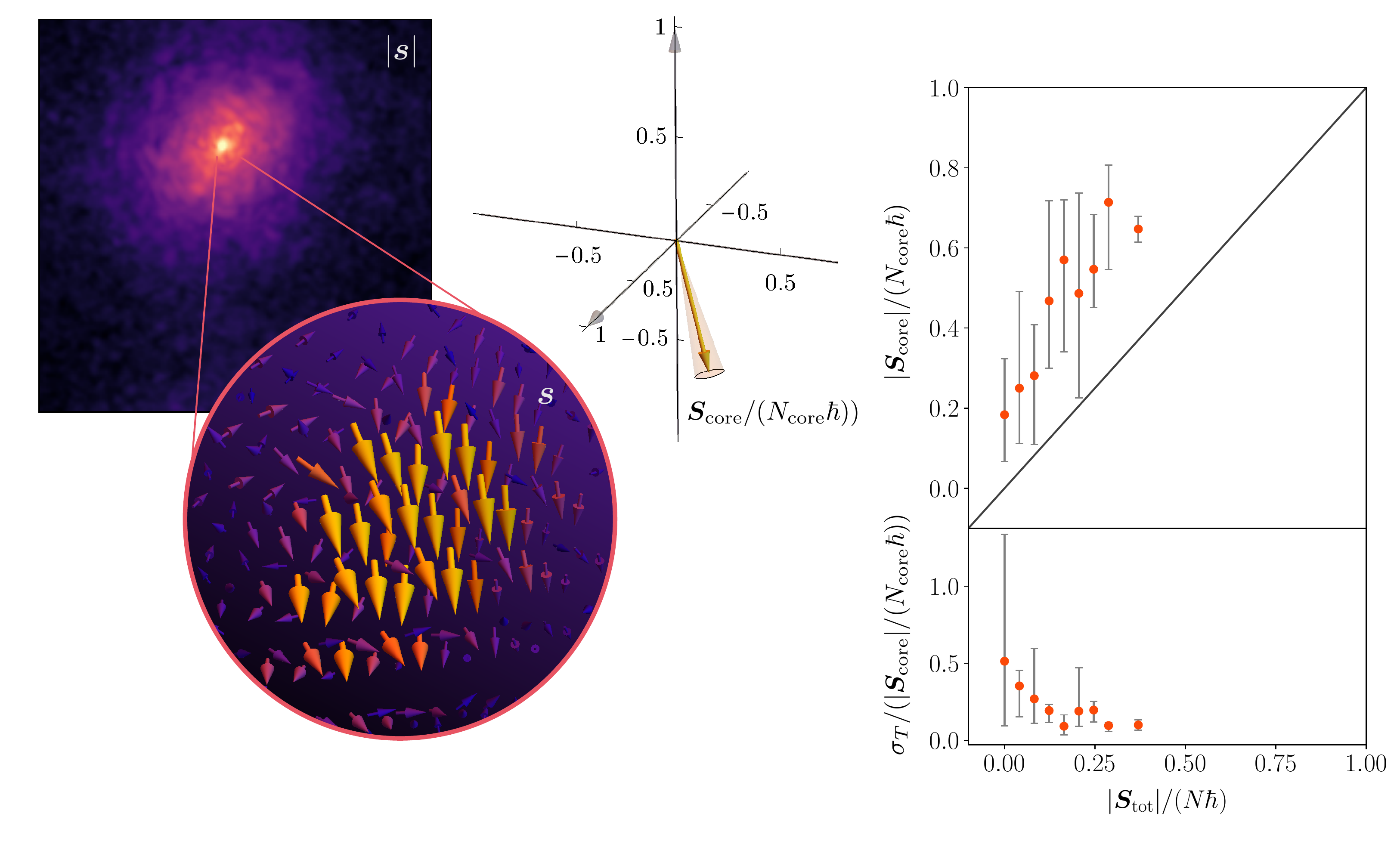}
    \caption{\small{The left panel shows the magnitude of the spin density at the end of the simulation, whereas the zoomed inset shows the vector spin density in the central core. The time-averaged spin vector per boson in the core (within $2r_c$) is shown in the middle, along with its typical precession around the mean  over a de-Broglie time scale. We take the smallness of the variation to be a sign that we have a soliton-like core in this cases, however, there can be larger variations in magnitude and direction of spin in other cases. The top panel of the rightmost plot shows a correlation between the initial spin per boson in our simulation (which is conserved) and the final spin per boson in the core. The red points are ensemble mean of the magnitude of the time-averaged vector spin in the core, where the ensemble consists of similar initial spin/boson simulations. The error bars show a 90\% confidence interval within this ensemble. The bottom panel shows the ensemble mean and standard deviation of the precession of the core spin. We caution that there might be a core, but not necessarily a soliton present at the centre in some of the cases. Note that a significant spin density in the core can be generated even at small initial values of the total spin.}}
    \label{fig:spinCompare}
\end{figure}

\subsection{Spin distribution}

We are interested in the intrinsic spin of the final configuration. To better understand how this intrinsic spin can be generated, we consider two classes of initial conditions. In one case, we start with initial conditions where each soliton has zero intrinsic spin, with $\bepsilon_i$ for each soliton a real random unit vector. In another class, each soliton has a random $\bepsilon_i$ complex vector (with unit norm), which leads to some residual non-zero intrinsic spin in the simulation volume (even though the spin directions for different solitons are not aligned). 

In Fig.~\ref{fig:spinCompare}, we show the final result of one such simulation (from the latter set), as well as the correlation between the initial spin in the box and the final spin in the core. At late times, we find that the final central core is fractionally polarized (with a mixture of circular and linear polarization) with a non-zero intrinsic spin. We notice a strong correlation between the initial spin in the box and the final spin of the core. For low initial spins, we find that the final spin in the core has a large variance as a function of time even at late times. This might be an indication that there is no settled soliton in the core for very low initial spins. For higher spins, there is a distinct direction of spin in the core, with relatively small variations with time (an example of which can be seen in Fig.~\ref{fig:spinCompare}). Wave interference effects in SDM create oscillations in the core radius, and induce a random walk of the core \cite{Li:2020ryg,Zagorac:2021qxq}. A similar analysis in VDM could shed light on the time dependence of spin in the core.

Since intrinsic spin for the entire simulation is conserved (separately from the  orbital angular momentum) in the non-relativistic limit, the halo carries the rest of the spin (with an opposite sign). 

\section{Observational Implications}\label{sec:end}

We discuss three application areas of VDM to astrophysical observations of interest, for possible future study.\footnote{For terrestrial experiments searching for light VDM, see for example, \cite{Chaudhuri:2014dla,Fedderke:2021aqo,Fedderke:2021rrm} and \cite{Antypas:2022asj} along with references therein. Also see \cite{Lisanti:2021vij,Nakatsuka:2022gaf} for the relevance of stochastic gradients in axion and dark photon fields. Related analysis can be repeated for the VDM case taking into account both VDM related interference patterns, solitons and their polarization.}

\subsection{Dark matter substructure and dynamical heating}
Density fluctuations resulting from wave interference in ultralight dark matter can dynamically heat the old stellar population in the Milky Way, thickening the scale height of its disk-like distribution \cite{Church:2019}. The heating rate $\mathcal{H}\propto (\delta\rho/\rho)^2\lambda_{\rm dB}^3\propto (\delta\rho/\rho)^2(\hbar/m)^3$ where $(\delta\rho/\rho)^2$ is the variance of fractional density fluctuations around the smooth density profile, and $\lambda_{\rm dB}\sim \hbar/m v$ is the characteristic size of the interference granules.\footnote{The heating rate:
    $\mathcal{H}\sim v^{-1}{(GM_{\rm gran})^2} n_{\rm gran}\ln \Lambda$, 
where $M_{\rm gran}\sim \delta\rho \lambda_{\rm dB}^3$ is the mass of a typical interference granule, $n_{\rm gran}\sim \lambda_{\rm dB}^{-3}$ is the number density of granules and $\Lambda$ is the usual Coulomb logarithm. We are essentially treating granules as scatterers which change the  velocity of test particles (stars). Using these approximations, we arrive at
    $\mathcal{H}\sim v^{-1}{(G\rho)^2}\left({\delta\rho}/{\rho}\right)^2\lambda_{\rm gran}^3\ln \Lambda$.
The smooth density profile $\rho$ and characteristic $v$ are assumed to be constrained observationally (though additional care might be needed here). For interference granules, $\delta\rho/\rho\sim 1$. See \cite{Church:2019} and  \cite{2022arXiv220305750D} for more details.} Using the scale height (and hence heating rate) as an observable for our galaxy, the authors in  \cite{Church:2019} argued for an SDM boson mass of $m \gtrsim 6\times10^{-23}~{\rm eV}$.

VDM would similarly lead to a dynamical heating of the old stellar disk. However, as we discussed in Section~\ref{sec:interference}, interference effects are reduced in VDM compared to SDM; $[(\delta\rho/\rho)^2]_{\rm vdm}/[(\delta\rho/\rho)^2]_{\rm sdm}\sim 1/3$. Furthermore, we expect the de Broglie scale in SDM and VDM to be the same for the same boson mass. These are confirmed by Fig.~\ref{fig:Correlation}. In case the boson masses are different, the size of the interference granules changes as $[\lambda_{\rm dB}]_{\rm vdm}/[\lambda_{\rm dB}]_{\rm sdm}\sim m_{\rm{sdm}}/m_{\rm vdm}$, while we still expect the same reduced interference in VDM compared to SDM. As a result, the constraint on $m_{\rm{vdm}}$ is relaxed by $3^{1/3}$ compared to the SDM case, ie. $m_{\rm vdm}>4.2\times 10^{-23}\,{\rm eV}$.\footnote{We thank Neal Dalal for pointing out an error in the previous version regarding the constraint on the vector boson mass.} A more careful analysis which takes into account the spectral information of the interference patterns and granules in VDM compared to SDM would be interesting to pursue. This analysis can be further generalized to spin-$s$ bosonic fields where the $[(\delta\rho/\rho)^2]_{ s}/[(\delta\rho/\rho)^2]_{ s=0}\sim 1/(2s+1)$.

A recent study has looked at constraining the SDM particle mass by considering the effect of dynamical heating on the stellar kinematics in ultra-faint dwarf galaxies
\cite{2022arXiv220305750D}, and found a tight constraint of $m \gtrsim 3\times10^{-19}~{\rm eV}$. We expect the constraint to be similarly relaxed by $\sim 3^{1/3}$ for VDM. Furthermore, as discussed in \cite{2022arXiv220305750D}, the property of the core might have a significant impact on this constraint (making it stronger). It would be interesting to consider the difference in the central core properties between VDM and SDM for constraining their respective masses.

\subsection{Dark matter cores in dwarf galaxies}
While standard CDM  predicts signature $r^{-1}$ cuspy centers of dark matter halos, at least without the presence of strong baryonic feedback, observational evidence strongly points to large low-density cores in dwarf galaxies (but see \cite{2020ApJ...904...45H}). 
Ultralight DM naturally points to a cored DM halo, and also predicts core sizes that are inversely proportional to mass. However, \cite{Deng:2018jjz,burkert2020fuzzy,2020ApJ...893...21S} point out that
a simple soliton profile given by its 1-parameter family solutions and a fixed boson mass has difficulty matching the diversity and scatter in observations of cores of halos with virial masses below $10^{11}~M_\odot$, despite the fact that 
the soliton profile can be matched well to dark-matter dominated dwarf spheroidal systems Sculptor and Fornax \citep{Marsh:2015wka}.
An outstanding issue is that when examining a broad sample of low mass galaxies, the inferred dark matter cores tend to have central densities $\rho_{\rm core}$ scaling inversely with core radius, $r_{\rm core}^{-1}$ \cite{Donato:2009ab,Karukes:2017kne}, 
while a soliton profile has a scaling $\rho_{\rm core}\propto r_{\rm core}^{-4}$ in both SDM and VDM.  An investigation of whether VDM may naturally introduce more scatter (and change in observed slope) into its core shape and the core-halo mass relation, as our idealized simulations suggest they may, is needed. In more realistic simulations, there could be additional scatter due to tidal stripping as suggested by \cite{Chan:2021bja} for the SDM case. It is conceivable that ultralight VDM might ameliorate the existing tensions with observational data further.

\subsection{DM Substructure and Lensing}
Multiply imaged quasars at high redshift probe dark matter substructure in halos;
in CDM theory populations of $10^9~M_\odot$ subhalos can alter the flux ratios of the images without significantly changing their positions
\citep{2001ApJ...563....9M}.
In VDM and SDM, while these theories predict a reduced population of subhalos, they also feature wave interference that causes additional lensing.
The projected density and its power spectra can be used to estimate the size of the effects of VDM and SDM (e.g. see plane wave perturbation analysis in \cite{1998MNRAS.295..587M}). The reduced interference effects in VDM compared to SDM  can change the lensing predictions, potentially probing the nature of the underlying field and changing constraints on the boson mass. Again, a more careful analysis is warranted for lensing due to substructure in VDM, including interference patterns and solitons.

\section{Summary}
\label{sec:summary}
Our goal in this paper was to understand whether the intrinsic spin of light dark matter has an impact on its small scale structure assuming only gravitational interactions. To this end, we have provided the first 3+1 dimensional simulation probing the small scale structure of vector dark matter (VDM) and compared the results with those of scalar dark matter (SDM). See Fig.~\ref{fig:snapCompare}. The key differences arise from the reduced interference of the fields in VDM compared to SDM.
\\ \\
\noindent{\bf Similarities}:
Starting with identical initial conditions in terms of their mass density (in the form of $N$ solitons), we found that in both cases we form an approximately spherically symmetric density configuration with a central core and a surrounding halo. The central core (after angular averaging) can be fit well by a single parameter (core radius $r_c$) soliton profile, and the surrounding halo eventually transitions to a power-law profile beyond $r\gtrsim 10 r_c$. The ratio of the mass of the central core to the total mass  can be expressed as a power law (with scatter, and different for VDM and SDM) of the initial energy of the system, $M_{\rm core}/M_{\rm tot}\propto \Xi^{\alpha}$.
\\ \\
\noindent{\bf Differences}\\
\noindent{\it Radial density profile}: For identical initial conditions, the central core is less dense and broader in VDM compared to SDM. If we normalize for the central density (or core radius), there still remains a distinct difference in shape of density distribution at the transition between the core and the halo: VDM has a smoother transition. This shape information is independent of the initial conditions. See Fig.~\ref{fig:radialDensity}.\\ \\
\noindent{\it Core-halo mass relation}: For VDM, we were able to analytically estimate $M_{\rm core}/M_{\rm tot}\propto\, \Xi^{\alpha}$ with  
$\alpha=-\log_2(\sqrt{f})$. The difference in the exponent betweem SDM and VDM can be attributed to different mass loss fractions, $f$, in individual binary soliton mergers (obtained numerically). 
Furthermore, there is $\sim 15\%$ more scatter of $M_{\rm core}/M_{\rm tot}$ in VDM simulations compared to SDM ones around the power-law fit. See Fig.~\ref{fig:CoreByXi}.\\ \\
\noindent{\it Interference}: The reduced interference in VDM compared to SDM leads to fewer extreme (underdense and overdense) regions in VDM compared to SDM. Once the region containing the core is removed, the amplitude of the two-point correlation function of the fractional density contrast  is smaller in VDM compared to SDM, while widths of the correlation functions are the same. Heuristically, the interference granules in VDM are of similar size, but less overdense than those in SDM. See Fig.~\ref{fig:rawDensities} and \ref{fig:2ptE}.
\\ \\
\noindent{\it Spin}: We initiated the study spin angular momentum during structure formation in VDM. See Fig.~\ref{fig:spinCompare}. While total spin is conserved, spin density undergoes a rich evolution. We found a positive correlation between the initial spin/boson in the simulation, and the final spin of the central core. Even for a small initial spin/boson, the final spin per boson in the core can be significant. The detailed dynamics of spin needs to be explored further. \\ \\

 \noindent{\bf Near Future Directions}: Much more remains to be done in the context of light, higher spin DM. The growing number of analyses of observational implications of ultralight SDM can be adapted for ultralight VDM, including for example effects on dynamical friction \cite{Hui:2016ltb,Wang:2021udl,Traykova:2021dua}, vortices \cite{Hui:2020hbq}, Ly$\alpha$ forest \cite{Irsic:2017yje,Nori:2018pka},  CMB and galaxy surveys \cite{Hlozek:2014lca,Lague:2021frh}. Again, we expect the difference in interference effects to hold the key to delineating effects of VDM and SDM. For a more careful analysis of observational constraints, and for additional handles on structure formation in VDM, the production mechanism for VDM  is relevant (e.g.~\cite{Graham:2015rva,Agrawal:2018vin,Co:2018lka,Long:2019lwl}). The production mechanisms can influence the initial shape of the power spectrum. A linear/quasi-linear analysis could then be used to provide initial conditions for a late time, large scale simulation of VDM (comparable to \cite{Schive:2014dra,Mocz:2019uyd,May:2021wwp} done for ultralight SDM). Finally, non-gravitational (electromagnetic) effects of the potentially large intrinsic spin of solitons and VDM cores will be worth exploring as a novel probe of the spin of DM in an astrophysical setting \cite{Jain:2021pnk}.
 \\ \\
\noindent{\bf Note:} Our manuscript  was submitted on the day \cite{Gorghetto:2022sue} appeared on the arXiv. In that paper, the authors investigate the formation of dark photon stars (vector solitons) in the early universe and their implication for VDM substructure. We will explore the connection between their results and ours in the future.

 \section{Acknowledgements}
 We would like to thank Neal Dalal for an insightful discussion on constraints on the boson mass, as well Ray Hagimoto, Siyang Ling, Andrew Long and HongYi Zhang for helpful conversations. MA and MJ are partly supported by a NASA grant 80NSSC20K0518. The numerical simulations used the NOTS cluster supported by the Center for Research Computing at Rice University.

\bibliographystyle{utphys}
\bibliography{reference}
\appendix

\newpage
\section{Appendix}\label{sec:sim}
\subsection{Simulation setup}
We simulate $\psi_i$ ($i=1,2,3$) governed by Eq.~\eqref{eq:SP_general} using a split-step Fourier method \citep{Mocz:2017wlg} (as was done for the scalar case, but now we have 3 components instead of 1). 
Our fields live on a discretized three dimensional grid with resolution $\Delta x$. This method has a temporal error of $\mathcal{O}(\Delta t^2)$, where $\Delta t$ is chosen by the Courant-Friedrichs-Lewy (CFL) condition \cite{Schwabe:2016rze}: $\Delta t \leq \max\Big[\frac{m}{6\hbar}\Delta x ^2 , \frac{\hbar}{m|\Phi|_{\max}}\Big]$, every iteration.

Quantities like mass and spin that are calculated with norm-preserving operations are conserved up to machine precision. We also monitor fractional energy loss $\Delta E/E_\mathrm{tot}$ as a function of time as a measure of our simulations fidelity. We have performed our simulations with $320^3$ grid points for $N$ soliton simulations, where we have observed $\Delta E/E_\mathrm{tot} \lesssim \mathcal{O}(10^{-5})$ for VDM and $\Delta E/E_\mathrm{tot} \lesssim \mathcal{O}(10^{-4})$ for SDM. We have used $256^3$ grid-points for two soliton simulations, where we have observed $\Delta E/E_\mathrm{tot}\lesssim \mathcal{O}(10^{-4})$ for both VDM and SDM.

\subsection{Estimating the final core mass}\label{sec:EstSolMass}
This calculation essentially follows the arguments in \cite{Du:2016zcv}.\footnote{Note that they refer to core as only the mass within $r_c$. This is $1/4$ of the mass of the soliton.}
Consider the case where we have $N$ solitons at time $t_0$, each of mass $M_{\rm sol}^{\rm i}=M_{\rm sol}^{(0)}$. At time $t_1$, they merge pair wise, to yield $N/2$ solitons, each with mass $M_{\rm sol}^{(1)}=2fM_{\rm sol}^{(0)}$. Here, $1-f$ is the fraction of the progenitor mass that does not contribute to the merged soliton. At time $t_2$, the $N/2$ solitons merge again to yield $N/4$ solitons, each with mass $M_{\rm sol}^{(2)}=2f M_{\rm sol}^{(1)}=(2f)^2 M_{\rm sol}^{(0)}$. This process continues until we have a single final soliton, this happen at time $t_n$ where $n$ is determined from $N/2^n=1$. That is, the merger process is complete after $n=\log_2(N)$ merger generations. With this, the final mass of the soliton $M_{\rm sol}^{\rm f}$ is
\begin{equation}
   M_{\rm sol}^{\rm f}/M_{\rm sol}^{\rm i}=(2f)^{\log_2(N)}.
\end{equation}

Of course this is all very crude, and mergers will likely not proceed with just equal mass solitons merging. Mergers with highly unequal mass are expected to not contribute increasing the mass of the merged soliton. This will reduce the effective $N$ to $\gamma N$. 

How can the vector and scalar cases be different? First, the $(1-f)$ fraction during merger might be different between VDM and SDM. We expect $f_{\rm v}<f_{\rm s}$. It is unclear to us whether the reduced interference might also lead to fewer mergers in VDM.

\subsection{Continuity equations for mass and spin density}
For our purposes, we were be particularly interested in following the mass density, and the spin density in the system. The equations of motion immediately lead to the following local conservation laws:
\begin{align}
&\partial_t\rho +\nabla\cdot \bm{\mathsf{j}}_\rho=0\,,\qquad\bm{\mathsf{j}}_\rho=i\frac{\hbar}{2}\left[\bPsi\nabla\bPsi^\dagger-\bPsi^\dagger\nabla\bPsi\right],\\
&\partial_t\bm{s} +\nabla\cdot \bm{\mathsf{j}}_{\rm s}=0\,,\qquad \bm{\mathsf{j}}_{\rm s} = \frac{\hbar^2}{2m}\left[\nabla\bPsi\times\bPsi^{\dagger} - \bPsi\times\nabla\bPsi^{\dagger}\right].
\end{align}
where $\bm{\mathsf{j}}_{\rho}$ and $\bm{\mathsf{j}}_{\rm s}$ are the mass current vector, and spin current tensor respectively. Similar equations can be written for other conserved quantities also.

\begin{figure}[t!]
    \centering
    \includegraphics[scale=0.4]{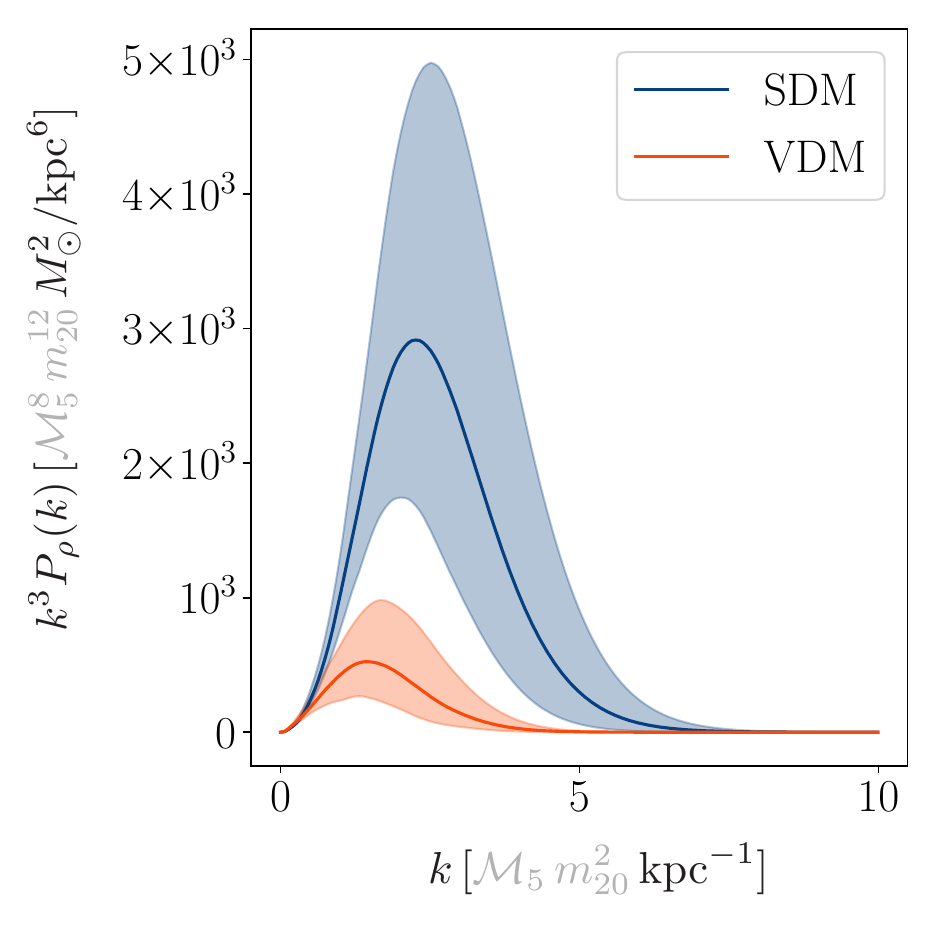}
    \caption{\small{$k^3$ scaled power spectrum of the density field for an ensemble of 10 simulations concentrated around $M=2.3\times10^5M_\odot\times \mathcal{M}_5$ with $N=21$ solitons each. The solid lines represent the mean, whereas the shaded bands include variation from different simulations.}}
    \label{fig:PS}
\end{figure}
\subsection{Power Spectrum}
\label{app:PS}
We construct the power spectrum of the density field: 
\begin{align}
    P_\rho(k)=\langle| \rho_{\bk}|^2\rangle\,,
\end{align}
with $\rho_{\bk}\equiv V^{-1/2}\int_V d^3 x\rho(\bx)e^{i\bk\cdot\bx}$ and $\langle \hdots\rangle$ representing an average over all $|\bk|=k$. Power spectra for an ensemble of 10 runs are shown in Fig.~\ref{fig:PS} (with $M=2.3\times10^5M_\odot\times \mathcal{M}_5$ and $N=21$ in each run).

From our simulations, we find that for VDM there is a characteristic peak at $k_{\rm peak}\approx 2.5\,\kpc^{-1} \times (\mathcal{M}_{5}m_{20}^2)$, whereas for SDM, the peak in the power spectrum is at $k_{\rm peak}\approx 3.2\,\kpc^{-1} \times (\mathcal{M}_{5}m_{20}^2)$. Note that for identical initial conditions, we find that typically the power in VDM is smaller, and peaks at smaller wave numbers compared to SDM. This is tightly correlated with the fact that the central soliton is less dense in VDM compared to SDM. 
\end{document}